\documentclass[twocolumn]{aastex631}
\usepackage{graphicx}
\usepackage{float}

\usepackage{lineno}
\usepackage{tablefootnote}
\DeclareUnicodeCharacter{2212}{\ensuremath{-}}

\newcommand{\lsim }{{\lower0.8ex\hbox{$\buildrel <\over\sim$}}}
\newcommand{\gsim }{{\lower0.8ex\hbox{$\buildrel >\over\sim$}}}
\newcommand{\Msun}{\ifmmode {M_{\odot}}\else${M_{\odot}}$\fi}
\newcommand{\Lsun}{\ifmmode {L_{\odot}}\else${L_{\odot}}$\fi}
\newcommand{\Rsun}{\ifmmode {R_{\odot}}\else${R_{\odot}}$\fi}

\shorttitle{NS ULXs and $L_R$/$L_X$}
\shortauthors{Panurach et al.}

\begin{document}

\title{Do Neutron Star Ultra-Luminous X-Ray Sources Masquerade as Intermediate Mass Black Holes in Radio and X-Ray?}

\correspondingauthor{Teresa Panurach}
\email{tpanurach@nsu.edu}

\author[0000-0001-8424-2848]{Teresa Panurach}
\affiliation{Center for Materials Research, Department of Physics, Norfolk State University, Norfolk VA 23504, USA \\}

\author[0000-0002-8532-4025]{Kristen C. Dage} \thanks{NASA Einstein Fellow}
\affiliation{Department of Physics and Astronomy, Wayne State University,  666 W Hancock St, Detroit, MI 48201, USA}
\affiliation{International Centre for Radio Astronomy Research -- Curtin University, GPO Box U1987, Perth, WA 6845, Australia}

\author[0000-0003-1814-8620]{Ryan Urquhart}
\affiliation{Center for Data Intensive and Time Domain Astronomy, Department of Physics and Astronomy, Michigan State University, East Lansing, MI 48824, USA \\}

\author[0000-0002-7092-0326]{Richard M. Plotkin}
\affiliation{Department of Physics, University of Nevada, Reno, NV 89557, USA}
\affiliation{Nevada Center for Astrophysics, University of Nevada, Las Vegas, NV 89154, USA}

\author[0000-0003-0040-3910]{Jeremiah D. Paul}
\affiliation{Department of Physics, University of Nevada, Reno, NV 89557, USA}

\author[0000-0003-2506-6041]{Arash Bahramian}\affiliation{International Centre for Radio Astronomy Research -- Curtin University, GPO Box U1987, Perth, WA 6845, Australia}

\author[0000-0002-4024-6967]{McKinley C. Brumback}
\affiliation{Department of Physics, Middlebury College, Middlebury, VT 05753, USA}

\author[0000-0002-2801-766X]{Timothy J. Galvin}
\affiliation{CSIRO Space \& Astronomy, PO Box 1130, Bentley WA 6102, Australia}

\author[0009-0004-4418-0645]{Isabella Molina}
\affiliation{Center for Data Intensive and Time Domain Astronomy, Department of Physics and Astronomy, Michigan State University, East Lansing, MI 48824, USA \\}

\author[0000-0003-2506-6041]{James C. A. Miller-Jones} \affiliation{International Centre for Radio Astronomy Research -- Curtin University, GPO Box U1987, Perth, WA 6845, Australia}

\author[0000-0002-5319-6620]{Payaswini Saikia}
\affiliation{Center for Astro, Particle and Planetary Physics, New York University Abu Dhabi, PO Box 129188, Abu Dhabi, UAE}

\begin{abstract}

Ultraluminous X-ray sources (ULXs) were once largely believed to be powered by super-Eddington accretion onto stellar-mass black holes, although in some rare cases, ULXs also serve as potential candidates for (sub-Eddington) intermediate mass black holes. However, a total of eight ULXs have now been confirmed to be powered by neutron stars, thanks to observed pulsations, and may act as contaminants for radio/X-ray selection of intermediate mass black holes. Here we present the first comprehensive radio study of seven known neutron star ULXs using new and archival data from the Karl G. Jansky Very Large Array and the Australia Telescope Compact Array, combined with the literature. Across this sample there is only one confident radio detection, from the Galactic neutron star ULX Swift J0243.6+6124.  The other six objects in our sample are extragalactic, and only one has coincident radio emission, which we conclude is most likely contamination from a background HII region. We conclude that with current facilities, neutron star ULXs do not produce significant enough radio emission to cause them to be misidentified as radio/X-ray selected intermediate mass black hole candidates. Thus, if background star formation has been properly considered, the current study indicates that a ULX with a compact radio counterpart is not likely to be a neutron star. 




\end{abstract}


\section{Introduction} \label{sec:intro}
Ultraluminous X-ray sources (ULXs) are bright, off-nuclear binary systems, and primarily extragalactic. These systems are known to host compact objects, and their observed X-ray luminosities ($L_x$) exceed the Eddington limit ($L_{Edd}$) for a 10$\Msun$ black hole, $L_x \sim 10^{39}$  erg s$^{-1}$. Because of their high luminosities, the population of ULXs were thought to be powered entirely by black holes, and some ULXs were also initially put forward as the missing class of intermediate mass black holes (IMBHs) \citep{Fabbiano89}. However, the discovery of pulsations from ULX M82 X-2 \citep{Bachetti14} disrupted the traditional ULX paradigm and presented just the first of many super Eddington neutron star ULX discoveries.  
It is now often suggested that most, if not all, ULXs, may be powered by neutron stars (\citealt[e.g.][]{2018ApJ...856..128W,King19}; see also reviews from \citealt{Kaaret17, Fabrika2021, 2023NewAR..9601672K, 2024arXiv240701768D}).

While pulsations are the best way to secure the identification of a neutron star primary, pulsations in ULXs are transient \citep{Bachetti14, Furst16, Israel17, Carpano18}. Because of a number of factors, such as geometry, thick outflows and disk warps, it is difficult to properly distinguish between potential black hole ULXs and neutron star ULX. That is, if pulsations are not observed, the object could still be a neutron star, with the site of pulsations obscured. As of yet, especially from X-rays alone, there is no metric to identify black hole ULXs and hence quantify if black holes represent a significant portion of the population of ULXs.


%




At low accretion rates, it has  previously been suggested that the ratio of radio to X-ray luminosity can distinguish black hole candidates from other classes of accreting compact objects \citep{Maccarone05a}, both in the  stellar ($\approx$10$M_\odot$; e.g., \citealt{Strader12, Chomiuk13, Miller-Jones15, Shishkovsky18, Zhao20, Tudor22})  and intermediate mass ranges ($\approx$10$^3$-10$^5$$\Msun$; e.g., \citealt{Nyland12, Webb12, Barrows19, Reines20, Paduano24}). The reason stems from black hole X-ray binaries (XRBs) in the hard X-ray spectral state displaying compact radio emission \citep{Fender01b, Corbel02}, which arises from a partially self-absorbed synchrotron jet \citep{Blandford79}.  In the hard state (when $L_X \lesssim 0.01 L_{\rm Edd}$), where black hole XRBs begin and end outbursts, they display a nonlinear correlation between their radio ($L_R$) and X-ray luminosities (which we will refer to as the radio/X-ray luminosity correlation; \citealt{Corbel03, Gallo03}).  While there is some overlap in the ratios of radio to X-ray luminosities ($L_R/L_X$) between black hole and neutron star XRBs, at a given $L_X$ black hole XRBs tend to have larger $L_R/L_X$ ratios by a factor of $\approx$20 on average \citep{Gallo18}.  Additionally, \citet{Merloni03} and \citet{Falcke04} demonstrated that by including a mass-normalization term, the radio/X-ray luminosity correlation can be extended to supermassive ($>$10$^6$ $M_\odot$) analogs of hard state black hole XRBs, through a non-linear correlation among $L_R$, $L_X$, and black hole mass $M_{\rm BH}$, which is termed the fundamental plane of black hole activity. Similar work has been done for neutron stars \citep[e.g.,][]{Gallo18,Gusinskaia20b, Panurach21}, however, objects studied in both X-ray and radio tend to bias towards black holes, as neutron star accretion dynamics are more complicated than black holes, and the periods of outburst are shorter.

If all ULXs are either super-Eddington stellar-mass black holes or neutron stars, they are in a significantly different accretion state than the sub-Eddington systems described in the previous paragraph.  As such, we do not expect ULXs with stellar-mass black holes to display the same luminosity correlations.  Nevertheless, we are inspired by the above to investigate whether, in the super-Eddington regime, there are radio signatures that might discriminate between black hole and neutron star ULXs.  Furthermore, joint radio and X-ray observations of ULXs are often used to distinguish IMBH candidates from stellar-mass XRBs, as well as for discovering active galactic nuclei (AGNs) in low-mass galaxies \citep[e.g.,][]{Mezcua13,Reines12, Webb12, Kim15,Yang16, Mezcua18}: the assumption being that a ULX would fall along the fundamental plane if powered by an IMBH or a sub-Eddington AGN, and it would fall off the fundamental plane if powered by a super-Eddington XRB. Characterizing the radio properties of known super-Eddington XRBs is therefore helpful for verifying the utility of using the fundamental plane, and for eventually quantifying the level to which super-Eddington XRBs (including neutron stars) could be contaminating AGN samples in low-mass galaxies.

Considering the above, in this paper we perform a  radio continuum survey of all known pulsating (i.e., neutron star) ULXs with radio coverage, which currently includes seven  of the eight known pulsating ULXs.   By focusing on pulsating ULXs, we can be confident  in the nature of the compact object. 
The majority of pulsating ULXs (M82 X-2, NGC 5907 ULX-1, M51 ULX-7, NGC 1313 X-2, NGC 300 ULX-1) have peak X-ray luminosities well in excess of $5\times10^{39}$ erg s$^{-1}$ \citep{Zampieri04, Kaaret06, Bachetti14, Israel17, Israel17s, Carpano18, vandenEijnden18, 2019MNRAS.488L..35S, RodriguezCastillo20, Furst23}. These sources, while variable, have been observed to be persistently luminous in X-rays over many years \citep{Furst23, Vasilopoulos20c, Hu21}. The other two sources often included as pulsating ULXs, located in the Milky Way and Magellanic Clouds respectively, are Swift J0243.6+6124 (hereafter Swift J0243) and  RX J0209.6-7427, which have peak X-ray luminosities reaching to $\sim 2 \times 10^{39}$ erg s$^{-1}$ in outburst \citep{Wilson-Hodge2018, 10.1093/mnras/staa1041}, and are transient, unlike the extragalactic ULXs in this sample.  We stress that the radio properties of some individual pulsating ULXs have already been studied and published \citep{Kaaret06, Mezcua13, Earnshaw16, vandenEijnden18}.   Our goal in this paper, despite the limited sample size, is to perform the first synthesis study on the radio properties of pulsating ULXs as a population.

This paper is structured as follows.  In Section \ref{sec:data}, we describe  new and archival radio data from the Karl G. Jansky Very Large Array (VLA, \citealt{2011ApJ...739L...1P}) and the Australia Telescope Compact Array (ATCA, \citealt{2011MNRAS.416..832W}). In Section \ref{sec:results}, we present our results from each of the seven sources. We then discuss the implications of our results by assessing how the radio properties change with the properties of the neutron star XRB, and how they impact our selection methods for massive black holes via X-ray/radio luminosity ratios. We summarize our findings in Section \ref{sec:summary}.

\section{Observational Data} \label{sec:data}
\subsection{Our Sample}
Our sample includes seven of the eight neutron star ULXs with observed pulsations. The ULXs included in this study are: M82 X-2, NGC 5907 ULX-1, M51 ULX-7, Swift J0243, NGC 300 X-1, NGC 1313 X-2, and NGC 7793 P13. The sample consists of persistent extragalactic systems, except for Swift J0243. We do not include RX J0209.6-7427 because there is no radio coverage of the source while it was in outburst.  Source parameters including R.A.\ and Dec., distances, and peak X-ray luminosities, are listed in Table \ref{tab:ULX}. 

\subsection{Radio Data}

The archival radio continuum data used in this study came from the VLA and the ATCA. The VLA data were taken primarily in C band (4.0 - 8.0 GHz) with the exception of NGC 5907 ULX-1, which has additional X band (8.0 - 12.0 GHz) data. ATCA observations cover the C and X bands simultaneously.

Three of the sources have ATCA data: NGC 1313 X-2 (Project Code: C2588, PI: Cseh), NGC 300 ULX-1 (Project Code: C3050 PI: Soria and Project Code: C3120 PI: Urquhart), and NGC 7792 P13 (Project Code: C3547 PI: Dage). All observations were taken with the extended 6km configuration. NGC 1313 X-2 and N7792 P13 were taken in two basebands centered at 5.5 and 9.0 GHz, each with 2.0 GHz of bandwidth. Data for NGC 1313 X-2 were taken on 2011 December 16 with a total of 21.16 h on the source and data for NGC 7792 P13 was taken on 2023 December 16 with 10.42 h on P13. These data were flagged and calibrated using standard procedures within the Common Astronomy Software Application (CASA) version 5.6.3 \citep{CASA2022} and imaged using the \verb|tclean| algorithm with Briggs weighting scheme and a robust setting of 1. Details of the observation and data processing of NGC 300  are described in \citet{Urquhart19}. 

The other sources: M82 X-2 (AL629, PI: Lang), NGC 5907 ULX-1 (12A-183, PI: Mezcua and 19A-148, PI: Middleton), M51 ULX-7 (12A-287, PI: Miller), and Swift J0243 (17B-406 and 17B-420, PI: van den Eijnden) were observed with the VLA in A or B configuration. 
We do not reanalyze the published observations of AL629 (M82 X-2), 12A-183 (NGC 5907 ULX-1), 17B-406 (Swift J0243) and 17B-420 (Swift J0243) and instead adopt values from \cite{Kaaret06,Mezcua13,vandenEijnden18}. The unpublished data of NGC 5907 ULX X-1 for program ID 19A-148 was observed four times with X band receivers (8.0 - 12.0 GHz) in A configuration with 41.4 min on target. The unpublished archival C band (4.0 - 8.0 GHz) data from M51 ULX-7 in 2012 were taken in B configuration with 10.6 min on the source. Each epoch was split into two 2.0 GHz baseband images centered at 9.0 and 11.0 GHz and 5.0 and 7.0 GHz, respectively. All flagging, calibration and imaging was done using {\sc CASA}ver.5.6.2 using the Briggs weighting scheme of robustness 1 to efficiently balance sensitivity and resolution of the image.

Observations that resulted in non-detections are reported in Table~\ref{tab:ULX} as $3\sigma$ upper limits, where $1\sigma$ denotes the root-mean-square noise in a source-free region surrounding the target location.


\subsection{X-ray Data}

While all of the pulsating ULXs have been well studied long-term with a number of X-ray observatories, we require X-ray luminosities in the 1-10 keV range for the purposes of the fundamental plane relation. Thus, we perform new extraction of one X-ray spectrum per source, and model it with the best fit model reported by the relevant literature, to calculate the 1-10 keV fluxes. 

The long-term monitoring of these sources has enabled a unique study of long-term variability in extragalactic X-ray sources. While sources like Swift J0243 and NGC 300 ULX-1 display transience (an outburst followed by a decay and subsequently not being detected above the Eddington limit), M82 X-2, NGC 1313 X-2, NGC 7793 P13, M51 ULX-7 and NGC 5907 ULX-1 all show long-term order of magnitude changes in flux \citep[e.g.,][and references therein]{2021A&A...652A.118R, 2023A&A...672A.140F, 2020MNRAS.495L.139T}. 

Given that our goal is to assess whether a neutron star ULX could confuse a search for an IMBH, we select one X-ray observation of each source when it is near its peak X-ray luminosity, extract the background subtracted source counts and model the spectrum. We note that this is already an issue with the hard-state condition for using the fundamental plane of radio/X-ray correlation to estimate mass; however, most extragalactic ULXs do not have of years of monitoring in X-ray and may have only been observed once above the Eddington limit\footnote{The discussion of variability and transience in ULXs when they are not well monitored is an ongoing discussion out of scope of this work but see \cite{2024MNRAS.52710185A, 2023ApJ...951...51B} and \cite{2021MNRAS.508.4008D} for a discussion of how the transient/variable nature of ULXs impacts our overall understanding of the population.}.

We further note that none of these X-ray observations are simultaneous with radio, and there are no instances where the archival X-ray data is strictly simultaneous with the archival (or new) radio data. Simultaneity is a requirement for using the fundamental plane. However, many ULXs are being discovered with archival data searches \citep[e.g.,][]{2022MNRAS.509.1587W}, and many ULX sources again do not have the luxury of wide-spread radio studies.

For the purposes of this study, it is illustrative enough to choose one bright X-ray observation out of the many available, but the short timescale of X-ray variability that is observed in most of these pulsating ULXs should serve as an additional caution for invoking the assumptions of the fundamental plane on a ULX where little is known. 

The choice of models for X-ray spectra is also an interesting discussion, as the shape and best-fit parameters of the spectral model can directly influence the measured X-ray flux. However, the ability to perform detailed modelling of the X-ray spectral shape is a function of the sensitivity of the instrument, as well as number of photons (i.e., both distance to the source and the intrinsic luminosity coming from the source). The pulsating ULXs in the present sample are both relatively nearby and extremely luminous, enabling very detailed modeling (and therefore a better understanding) of their X-ray spectral states via observatories such as NuSTAR and XMM-Newton, but many other ULXs do not share this luxury.

\begin{deluxetable*}{lccccccc} 

\label{tab:ULX}
\tabletypesize{\footnotesize}
\tablewidth{0pt} 
\tablecaption{NS ULX Radio and X-ray luminosities. Program IDs marked with $\alpha$ means that we adopt the published radio luminosity from \cite{Kaaret06}, $\beta$ is the published radio luminosity from \cite{Mezcua13}, and $\gamma$ are radio luminosities measured and published by \cite{vandenEijnden18}.}
\tablehead{Source & R.A. & Dec. & Distance  & Radio Program & Observation   & Flux Density &  Peak X-ray Luminosity\\
 & (h:m:s) & ($^{\circ}$:\arcmin:\arcsec)   & (Mpc) & ID & Date & ($\mu$Jy) & (1-10 keV erg s$^{-1}$)}
\startdata
M82 X-2 & 09:55:51.04 & 69:40:45.49 & 3.6 & VLA AL629$^{\alpha}$ & 2005 Jan 29 & 2.9 $\pm$ 0.1 mJy (8.5 GHz) & 9.0 $\times 10^{39}$\\
NGC 5907 ULX-1 & 15:15:58.62 & 56:18:10.3 & 17.1 &  VLA 12A-183$^{\beta}$& 2012 May 30 & $<$12.9 (5.0 GHz) & 7.6 $\times$ $10^{40}$\\
& &  & & VLA 19A-148 & 2019 March 05 & $<$18.3 (9.0 GHz) \\
& & & &  & 2019 April 30 & $<$15.3\\
& & & & & 2019 May 30 &  $<$29.4\\
& & & &  & 2019 July 01 &  $<$22.2\\
M51 ULX-7 & 13:30:01.098 & 47:13:42.33 & 8.58 & VLA 12A-287& 2012 June 02 & $<$53.4 (6.0 GHz) &  4.0 $\times$ $10^{39}$\\
Swift J0243 & 02:43:40.43 & 61:26:03.76 & $5.2\times10^{-3}$ &  VLA 17B-406$^{\gamma}$& 2017 Oct 10  & $<$12 (6.0 GHz) & 8.7 $\times 10^{38}$\\
& & &  &  & 2017 Nov 08 &77.1 $\pm$ 4.2\\
& & & & VLA 17B-420$^{\gamma}$ & 2017 Nov 15&92.6 $\pm$ 3.8\\
& & & & & 2017 Nov 21 &63.4 $\pm$ 4.3\\
& & & & & 2017 Nov 23&55.3 $\pm$ 4.4\\
& & & & & 2017 Nov 28&34.8 $\pm$ 4.0\\
& & & & & 2017 Dec 02&24.7 $\pm$ 4.5\\
& & & & & 2018 Jan 09&21.3 $\pm$ 4.0\\
NGC 300 ULX-1 & 00:55:04.86 & \-37:41:43.7 & 1.88 & ATCA C3050 &  2015 Oct 21-23 & $<$14.4 (5.0 GHz) &  2.4 $\times$ $10^{39}$\\
NGC 1313 X-2 & 03:18:22.18 & -66:36:03.3 & 3.7 & ATCA C2588 & 2011 Dec 16 & $<$124.5 (5.0 GHz) & 6.0 $\times$ $10^{39}$\\
NGC 7793 P13 & 23:57:50.9 & −32:37:26.6 & 3.83 & ATCA C3547 & 2023 Dec 16&$<$36.2 (5.0 GHz)& 3.7 $\times 10^{39}$
\enddata
\end{deluxetable*}

With these caveats in mind, we perform the X-ray analysis as follows:  

All of the sources where we use an observation from Swift/XRT were extracted via the build products page hosted by Swift\footnote{\url{https://www.swift.ac.uk/}} \citep{2009MNRAS.397.1177E}, using HEASOFT v6.32.  All of the \textit{Chandra} \citep{2000SPIE.4012....2W} observations we used were reprocessed with \textsc{chandra$\textunderscore$repro} using \textsc{ciao} version 4.15 \citep{2006SPIE.6270E..1VF} and HEASOFT v6.31.1. All of the background subtracted source spectra were extracted with \textsc{specextract} and binned by 20 counts. The extraction source regions are discussed on a case-by-case basis below, and the background for all sources was extracted by placing 6 of the same size regions around the source region, making small shifts to avoid any nearby point sources while still sampling the nearby background. All spectra were fitted with \textsc{xspec} 12.13.0c \citep{Arnaud96}, abundances from \cite{Wilm}, and photo-electric cross sections from \cite{1996ApJ...465..487V}. For all sources, we obtained line-of-sight neutral hydrogen density ({$N_{\text{H}}$}) from \textit{Chandra} \footnote{\url{https://cxc.harvard.edu/toolkit/colden.jsp}} and performed fits with \texttt{tbabs} fixed to this value. For all, we ignored any  bad bins. For the \textit{Chandra} observations, we fit within the 0.5-8.0 keV range, binned by 20 with $\chi^s$ statistics. As the Swift observations had total source counts an order of magnitude lower than the \textit{Chandra} observations, due to lower instrument sensitivity and overall lower exposure times, we fit these with W-stat \citep{1979ApJ...228..939C}, binned by 1, and used Anderson Darling as the test statistic. We fit the \textit{Swift} photon-counting (PC)  mode  observations in the 0.3-10 keV regime, and the \textit{Swift} windowed timing (WT) mode in the 0.7-10 keV range. All errors reported are at the 90\% interval.

For M82 X-2, we selected ObsID 16580 (PI: Margutti, 50 ks, 2014-02-03, ACIS-S) based on a previous study by \cite{2016ApJ...816...60B}.  We used a source extraction region of 1.5 arcsecond. As discussed by \cite{2016ApJ...816...60B}, this was an on-axis observation which was affected by pileup, and the overall fit statistics indicated that a blackbody disk \citep{1984PASJ...36..741M} was the best fit model for this source. We performed a fit with the model \texttt{pileup*tbabs*tbabs*diskbb}, keeping the first \texttt{tbabs} component fixed to the line-of-sight {$N_{\text{H}}$} as discussed above, and leaving the second column density free to measure intrinsic absorption. To obtain the 1-10 keV flux, we used \texttt{cflux}, removing the pileup model and freezing the fit parameters. Our fit values are reported in Table 2.  
The fit statistic is 239.81 with 271 bins and a null hypothesis probability of 8.83$\times 10^{-1}$ with 267 degrees of freedom. Within errors, our fit parameters are consistent with those of  \cite{2016ApJ...816...60B}.

For M51 ULX-7, the source was marginally off-axis and we extracted with a 5 arcsecond region. We selected ObsID 13814 (2012-09-20, 190 ks, PI:Kuntz, ACIS-S).  The source count rate was not high enough to have an issue with pileup. \cite{Earnshaw16} report that this source is best fit with a single power-law model. We initially attempted a second free absorbing column, but the best fit value was consistent with zero, so we omitted it and only fit \texttt{tbabs*powerlaw}, with \texttt{tbabs} frozen to the line-of-sight value.

Our fit values are reported in Table 2. The fit statistic was 220.16/205, and a null hypothesis probability of 1.81 $\times 10^{-1}$ with 202 degrees of freedom.
This is consistent with \cite{Earnshaw16}'s results.

NGC 1313 X-2 was off axis and we used a 10 arcsecond extraction region.  We used ObsID 2950 (2002-10-13, 20 ks, PI: Murray, ACIS-S). Due to the high source count rate, we modeled pileup with the $\alpha$ parameter frozen to 0.5 \footnote{\url{https://cxc.harvard.edu/ciao/download/doc/pileup_abc.pdf}}.
Fitting with an absorbed \texttt{diskbb} model provided better fit statistics (216.34/208), than a power-law   (237.4/208). Our fit values are reported in Table 2. 
Our fit parameters are consistent with \cite{2008AIPC.1054...39K}.

We modeled NGC 7793 P13 similarly to NGC 1313 X-2, which was also off axis and affected by pileup. We used ObsID 3954 (2003-09-06, PI: Pannuti, 50 ks, ACIS-S). We used an extraction region of 7 arcsec. Once again $\alpha$ was frozen to 0.5.  The fit statistics for a disk (355.41/221 bins) were worse than a power-law (244.1/221 bins), so we fit \texttt{pileup*tbabs*tbabs*powerlaw}. The fit values are reported in Table 2. 
We note this fit is consistent with \cite{Israel16a}.

For NGC 300 ULX-1, we used ObsID 00049834010, which was observed by Swift in PC mode for 5194 seconds on 2017-04-22 (PI: Binder). \cite{2018ApJ...863..141B} report 1-10 keV fluxes, and fit 11 Swift observations simultaneously. We selected the longest observation, ID 00049834010, and fit it with a simpler absorbed power-law model. We did not use a second unfixed absorbing column, as it was consistent with zero, and we fixed our line-of-sight value to $3.08 \times 10^{22}$ cm$^2$.  For comparison, we also fit an absorbed blackbody disk model, and the fits are reported in Table 2. The Anderson Darling fit statistics for the power-law model are -6.96 using 325 bins, and for the diskbb model are -6.05 using 325 bins.  
 We also note that \cite{2018ApJ...863..141B} found that the X-ray luminosity was consistent across their 11 observations, and it is also consistent, within uncertainties,  with the ones we have obtained by modeling with simpler spectra. 

For NGC 5907, we used ObsID 00032764009 (PI Pintore), taken on 2018-02-14 using Swift with photon counter (PC) mode for a 2266 second exposure. \cite{Israel16b} fit their Swift observations with tbabs*bknpowerlaw model, with absorption, energy cutoff and both power-laws fixed. When we fit with their fixed parameter model and cflux, we recover a 1-10 keV flux of 2.19 $\pm 0.48\times 10^{-12}$. For the simpler model of \texttt{tbabs*tbabs*powerlaw}, the Anderson Darling test was -5.44, and we report the fits in Table 2.  

\begin{table*}[]
\label{table:uglyfits}
\caption{Best fit parameters and fluxes for ULXs fit with \texttt{tbabs*tbabs*diskbb} or \texttt{tbabs*tbabs*powerlaw}. }
\begin{tabular}{lllllll}
Source         & $\alpha$                & $N_{\text{H}}$ (cm$^{-2}$) & kT (keV)                & $\Gamma$        & Norm.                              & \begin{tabular}[c]{@{}l@{}}Flux  erg/s/cm$^2$\end{tabular} \\
M82 X-2        & 0.35 $^{+0.10}_{-0.05}$ & 2.85 $\pm$ 0.20            & 2.48 $^{+0.36}_{-0.28}$ & -               & 2.4 $\pm$ 0.2 $\times$ $10^{-2}$ & 5.8 $\pm$ 0.1 $\times 10^{-12}$                           \\
M51 ULX-7      & -                       & -                          & -                       & 1.5 $\pm 0.1$ & 6.7 $\pm 0.4 \times 10^{-5}$       & 4.5  $\pm 0.1 \times 10^{-13}$                               \\
NGC 1313 X-2   & 0.5                     & 0.12 $\pm 0.02$            & 1.6 $\pm 0.1$          & -               & 0.5 $\pm 0.1$                     & 3.7 $\pm 0.1 \times 10^{-12}$                                \\
NGC 7793 P13   & 0.5                     & 0.06 $\pm$ 0.02            & -                       & 1.2 $\pm$ 0.1  & 3.9 $\pm 0.2\times 10^{-3}$        & 2.1 $\pm 0.1 \times 10^{-12}$                              \\
NGC 300 ULX-1  & -                       & -                          & -                       & 1.2 $\pm $ 0.1  & 6.3 $\pm 0.5 \times 10^{-4}$       & 6.8 $\pm 1.0 \times 10^{-12}$                              \\
NGC 300 ULX-1  & -                       & -                          & $2.3 \pm 0.4$           & -               & 1.2 $\pm 0.4 \times 10^{-2}$       & 5.8 $ \pm 0.8 \times 10^{-12}$                             \\
NGC 5907 ULX-1 & -                       & 1.16 $^{+1.25}_{-0.89}$    & -                       & 1.8 $\pm 0.9$   & 6.1 $\pm 4.1 \times 10^{-4}$       & 2.9 $\pm 1.0 \times 10^{-12}$                              \\
\end{tabular}
\end{table*}

Swift J0243 was observed seven times in outburst, in WT mode (PIs: Heinz, Kennea, Wolff).  \cite{vandenEijnden18} fit \textsc{tbabs*bbodyrad*powerlaw}, cutting off energies below 0.7 keV to all observations except 00010467008, which was fit with \textsc{tbabs*(powerlaw}. Because we are only interested in the 1-10 keV fluxes, and \cite{vandenEijnden18} have already performed detailed X-ray spectroscopy, after ensuring that we recover their best-fit model parameters, we use \textsc{cflux} to obtain the 1-10 keV flux, which are presented in Table 3.
\begin{table*}[]
\label{table:xrayswift}
\caption{Observations and 1.0-10 keV fluxes for Swift J0243.}
\begin{tabular}{llll}
ObsID       & Exp. Length (sec) & Obs. Date  & Flux ($\times 10^{-8}$ erg/s/cm$^2$) \\
00010336007 & 1931              & 2017-10-10 & 0.94 $\pm$ 0.03                                                          \\
00010336022 & 1054              & 2017-11-09 & 26.89 $\pm$ 0.01                                                         \\
00010336025 & 1019              & 2017-11-15 & 16.31 $\pm$ 0.01                                                         \\
00010336031 & 989               & 2017-11-27 & 7.24$\pm$ 0.01                                                          \\
00010336033 & 1029              & 2017-12-01 & 4.82$\pm$ 0.01                                                          \\
00010467007 & 899               & 2018-01-02 & 1.60$\pm$ 0.01                                                          \\
00010467008 & 1034              & 2018-01-13 & 1.10$\pm$ 0.01                                                         
\end{tabular}
\end{table*}

\cite{vandenEijnden18} also interpolated X-ray fluxes for radio observations taken at 2017-11-21 and 2017-11-23 with log-linear interpolation. We also performed log-linear interpolation and for 2017-11-21, we interpolate the 1-10 keV X-ray flux to be 1.01 $\times 10^{-7}$ erg/s/cm$^2$ and 2017-11-23, we interpolate a 1-10 keV X-ray flux of 8.87 $\times 10^{-8}$ erg/s/cm$^2$. 



\section{Results and Discussion} \label{sec:results}

We describe individual sources and their radio/X-ray properties below.  To briefly summarize, the only radio-detected source is the Galactic ULX Swift J0243 \citep{vandenEijnden18}.  A radio counterpart is also detected from M82 X-2, but this likely arises from an \ion{H}{2} region and not the ULX (see Section \ref{sec:res:m82}).

Before discussing each source, we first comment that radio counterparts to ULXs are relatively rare.  When radio emission is detected, it usually appears as a `radio bubble', which is an extended nebular feature that  emits optically thin synchrotron radiation, likely associated with an active or past outflow shocking the nearby interstellar medium \citep{2010Natur.466..209P,2010MNRAS.409..541S, 2014MNRAS.439L...1C, 2015MNRAS.452...24C, Urquhart19, Anderson19}. Typical bubble sizes range from 30 to 300 parsec \citep{Berghea20}. No pulsating ULX (including Swift J0243) has yet to be linked to radio bubbles. Radio emission from jets is also possible but so far has only been seen in a handful of ULXs \citep{Middleton2013,Mezcua13,Cseh2015, Yang16}.


\subsection{M82 X-2}
\label{sec:res:m82}
M82 X-2 was the first detected neutron star ULX, confirmed through X-ray pulsed emission from \emph{NuSTAR} in 2014 \citep{Bachetti14}. The source has an orbital period $\sim$2.5 d \citep{Bachetti22}. The predicted magnetic field strength is B  $\approx$ 10$^{12}$G \citep{Bachetti14, Eksi15, Tsygankov16} but may be as low as $<$10$^{9}$G \citep{Kluzniak15}. The peak X-ray luminosity of the source is 2 $\times$ $10^{40}$ erg s$^{-1}$ in the 0.3--10 keV band \citep{Bachetti14}. 

Previous radio flux density measurements were made using the VLA in 2005 (VLA/AL629, PI: Lang), and 
\citet{Kaaret06} reported a significant ($\sim$ 1 mJy) detection near the position of the ULX, and we adopt this value for subsequent analysis in this paper. As seen in their Figure 8, the ULX source is quite offset from the center of the radio contours. We also note that  a source was detected at the position of the ULX in an eMERLIN survey \citep{2013MNRAS.431.1107G}, and classified as an HII region (source 16 or 42.20+59.1 in their Table 2). If we take the detection at face value, it corresponds to a radio luminosity of 1.3 $\times 10^{35}$ erg s$^{-1}$, which is unreasonably luminous to expect from an XRB, but typical for an HII region \citep{2013MNRAS.431.1107G}. While M82 X-2 is located in an extremely dusty part of the host galaxy, James Webb Space Telescope observations may be able to provide better insight to the ULX's environment. Regardless, this underscores the impact that background star formation can have on radio measurements of XRBs.

\subsection{NGC 5907 ULX-1}
 NGC 5907 ULX-1 was observed in radio by \cite{Mezcua13}, who published an upper-limit ($<0.020$~mJy) in a 5.0 GHz VLA observation (VLA/12A-183, PI: Mezcua). We adopt this upper limit value for this paper. They used the fundamental plane of black hole activity to infer that the compact object is a black hole with a mass upper limit of $3 \times 10^{3}$ $\Msun$. More recently, X-ray pulsations have been detected with a 1.43s spin period and a peak X-ray luminosity of 5 $\times$ $10^{40}$ erg s$^{-1}$ \citep{Israel17s}, clearly ruling out a black hole scenario. 

\subsection{M51 ULX-7}
M51 has been subjected to multiple observational campaigns in various wavelengths due to its high star formation rate and large number of X-ray sources \citep{Kilgard05}. M51 ULX-7 is a pulsar ULX system at a distance of 8.6 $\pm$ 0.1 Mpc \citep{McQuinn16}. The neutron star has a 2.8s spin period and an orbital period of $\sim$ 2 d \citep{RodriguezCastillo20}. The system has a peak X-ray luminosity of $\sim$10$^{40}$ erg s$^{-1}$ \citep{RodriguezCastillo20} and was discovered to exhibit superorbital modulations from its \emph{Swift}/XRT lightcurves at a period of 38 days \citep{Brightman20,Vasilopoulos20c}. 

Previous radio observations of M51 ULX-7 are limited. 
Observations of ULX-7 %
have been published in \citet{Earnshaw16} using L band (1.0 - 2.0 GHz) data from VLA/11A-142 (PI: Hughes). The source was not detected during this observation, measuring an upper limit of $<87 \mu$Jy at 1.5 GHz. In this study, we use archival C band (4.0 - 8.0 GHz) data from the VLA/12A-287 (PI: Miller) on 2012 June 02, measuring an upper limit of $<53 \mu$Jy at 6.0 GHz. 

\subsection{Swift J0243}
Swift J0243 is one of the few super-Eddington sources in our Galaxy. There have been many previously reported distance values from $\sim2.5 - 6$kpc \citep{Bikmaev17,vandenEijnden18,Zhang19,Reig20}, and most recently from the latest Gaia DR3 catalog, $5.5_{-0.3}^{+0.4}$kpc \citep{Gaia22}. For this study we use a distance of 5.2 kpc, adopting the lower limit similarly to \citet{vandenEijnden18}. Swift J0243 was first discovered by Swift's Burst Alert Telescope (\emph{Swift}/BAT, \cite{2004ApJ...611.1005G}) in 2007 through a giant outburst \citep{Cenko17} and was found to exhibit X-ray pulsations with a period of 9.86s \citep{Kennea17,Bahramian17atell,Wilson-Hodge2018}, establishing the compact primary to be an accreting neutron star. Later optical spectroscopic observations of Swift J0243 showed it to have a Be counterpart, making it a part of a special class of high-mass X-ray binaries, Be/X-ray sources \citep{Reig20} with a strong magnetic field \citep[$B \geq 10^{12}$\,G;][]{Doroshenko18}. 

Swift J0243 is one of the few ULX sources with  multiple high-quality quasi-simultaneous radio and X-ray observations.  \cite{vandenEijnden18} followed the source from outburst to quiescence over eight C band (4.0 - 8.0 GHz) VLA (17B-406 and 17B-420, PI: van den Eijnden) and seven \emph{Swift/XRT} observations.  They observed radio and X-ray emission that was correlated, but seen at a much lower level than neutron stars with low magnetic fields. Thus, while high magnetic fields apparently do not inhibit jet launching in ULXs, they cause the jets to be much fainter in radio.  From \cite{vandenEijnden18}, we adopt 5.0 GHz radio andmeasure 1--10\,keV X-ray luminosities to place these epochs on the radio/correlation plane in Figure \ref{fig:lrlx}. 

\subsection{NGC 300 ULX-1}
NGC 300 ULX-1 was initially thought to be a supernova, SN2010da \citep{Khan10} and later was categorized as a Wolf Rayet/black hole XRB system \citep{Carpano07} until it was discovered to have X-ray pulsations through \emph{NuSTAR} \citep{Harrison13} and \emph{XMM-Newton} \citep{2001A&A...365L..18S} with a strong periodic modulation (31.6s) in the X-ray flux \citep{Carpano18}. It reaches a peak X-ray luminosity of $5 \times 10^{39}$ erg s$^{-1}$ \citep{Carpano18}. 

For our analysis, we reduced archival ATCA observations from 2014, with the pointing center 2\arcsec\ from the NGC 300 ULX-1 system (Project Code: C3050 PI: Soria and Project Code: C3120 PI: Urquhart). We measure a flux density upper limit of $<$11$\mu$Jy at 5.0 GHz, and we adopt the X-ray flux shown in Table \ref{tab:ULX}.

\subsection{NGC 1313 X-2}
NGC 1313 X-2 has a 1.5s spin period and a peak X-ray luminosity of $2 \times 10^{40}$ erg s$^{-1}$ \citep{2019MNRAS.488L..35S}. While it was observed by ATCA in 2011 (Project Code: C2588, PI: Cseh), the center of the field pointed approximately 30 arcseconds from the pulsating ULX source. The ULX is far enough off the pointing center that it is more challenging to obtain a meaningful 3$\sigma$ upper limit with our best constraint being $<$124$\mu$Jy at 5.0 GHz. 

\subsection{NGC 7793 P13}
NGC 7793 P13 was first identified in 2003 as an X-ray source in a Chandra X-ray Telescope survey of NGC 7793 \citep{Pannuti06}. Previous Swift/XRT monitoring of P13 in 2010 showed a range of L$_{\rm x}$ $\sim 3 - 5  \times 10^{39}$ erg s$^{-1}$ \citep{Motch14} and later found to peak at luminosities of $10^{40}$ erg s$^{-1}$ \citep{Israel17}. It is the fastest pulsating ULX, with a spin period of 0.42 seconds. It was observed with ATCA in Dec 2023 (ATCA/ C3547, PI: Dage) with an 3$\sigma$ upper limit radio flux density of $<$36$\mu$Jy at 5.0 GHz.



\subsection{Do Neutron Star ULXs Contaminate the Fundamental Plane?}
\label{sec:disc:fp}

One objective of this study is to identify whether ULXs that are known neutron stars could act as contaminants to black hole candidates selected through the radio/X-ray luminosity correlation and/or the fundamental plane. I.e., if their identification was not secure as neutron stars, could they be mistaken as massive black hole candidates based off their X-ray and radio emission? 

First, a strong word of caution is that the radio/X-ray luminosity correlation, and by extension the fundamental plane, is derived from hard-state XRBs and their supermassive analogs \citep{Merloni03, Gallo03,Falcke04, Coriat11,Gallo18,Gultekin19,Shen2020}. ULXs have their own distinct spectral states  \citep{Gladstone2009, Sutton13, Urquhart16}. 
 Unless a ULX is a sub-Eddington IMBH, it is not expected to fit along the radio/X-ray luminosity correlation.  
As such, we stress that there is no physical motivation to place a known pulsating ULX onto the radio/X-ray luminosity correlation or onto the fundamental plane.  Rather, the exercise here is to assess the level to which pulsating ULXs could masquerade as black holes during joint radio/X-ray studies, should one not be aware of the nature of the compact object.  
 We also note that many of the non-transient ULX sources in our sample are highly variable, on the scale of an order of magnitude in X-ray flux \citep[e.g.,][]{Furst21}, which would introduce extra scatter in radio/X-ray luminosity correlations and in the fundamental plane. Ideally, radio and X-ray  observations would be taken taken simultaneously, which was unfortunately not feasible for most objects in our sample, with the exception of \citet{vandenEijnden18}'s monitoring campaign on Swift J0243.

\begin{figure*}[t]
\epsscale{1.15}
\plottwo{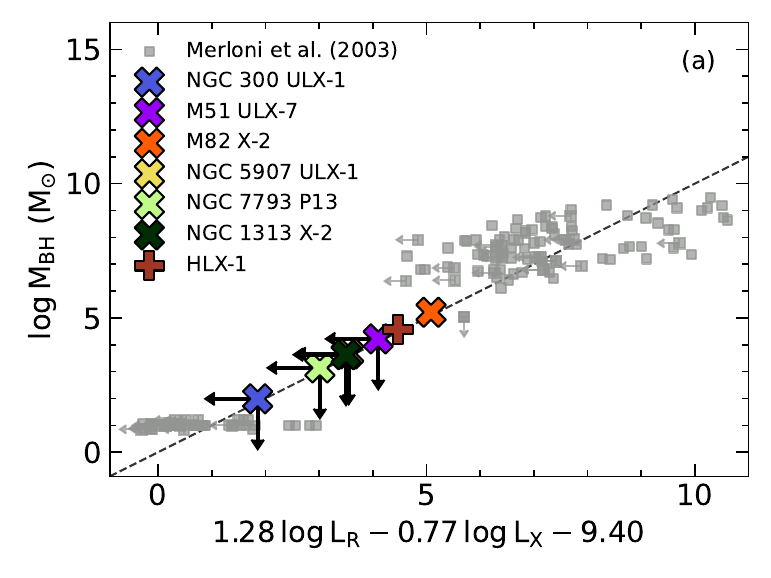}{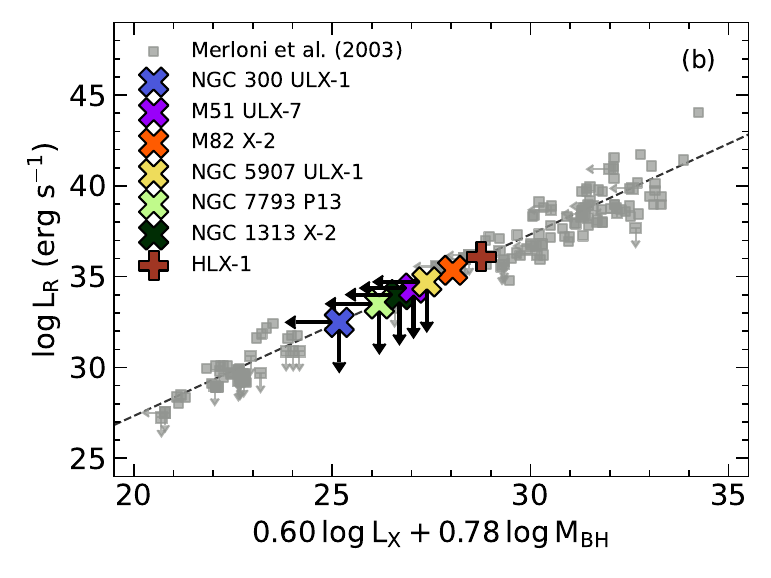}
\caption{Our sample of extragalactic neutron star ULXs (shown by cross symbols) projected on to the fundamental plane, showing the result if one were to naively estimate ``black hole" masses via the radio and X-ray luminosities/limits (here we have used Eq.~15 of \citealt{Merloni03}). Grey symbols show XRBs and AGN from \cite{Merloni03}, and the dashed line shows their best-fit function. Since it is common for studies to employ joint radio and X-ray observations to search for AGNs in low-mass galaxies, this figure illustrates the parameter space where neutron star ULXs could potentially masquerade as IMBH/low-mass AGNs. (a): A projection of the fundamental plane with black hole mass as the dependent variable. The radio and X-ray luminosities/limits of our sample have led to erroneous mass estimates or upper limits of $\sim 10^2$$-$$10^5 \Msun$. (b): The same data projected onto the typical view of the fundamental plane. Note: The radio counterpart to M82 X-2 is likely from a coincident H II region, but we depict it as a radio detection here to illustrate how such a mis-identification would incorrectly imply a $\approx 10^5\,M_\odot$ IMBH.  The Galactic source Swift J0243 is not included in this figure, as the fundamental plane correctly predicts this source cannot be a black hole (see Section 3.8).  For comparison, we also include HLX-1 (as a plus symbol) as measured in X-ray and radio by \cite{Cseh2015}. }
\label{fig:fp}
\end{figure*}

With the above caveats in mind, if one were to blindly place our sample onto the fundamental plane (Figure~\ref{fig:fp}, using masses estimated via Eq.~15 of \citealt{Merloni03}), the results would be largely misleading.  The only exception is the Galactic ULX Swift J0243, where the fundamental plane  predicts a mass of $10^{-4}~\Msun$, which is clearly inconsistent with a black hole. At the other extreme, an egregious  error  could  occur if one were to not recognize the radio counterpart to M82 X-2 as a background \ion{H}{2} region \citep{2013MNRAS.431.1107G}, in which case the fundamental plane would suggest a $\sim10^5 \Msun$ IMBH.  One could envision this type of mistake happening if a system like M82 X-2 were part of a larger sample derived from cross-matching X-ray and radio catalogs (a scenario where investigating mis-identifications on an individual basis is often not feasible).  

For the other five ULXs in our sample with radio non-detections, the fundamental plane suggests black holes with mass upper limits ranging from $\lesssim$50--10,000 $\Msun$.  The severity of this (misleading) calculation on the final interpretation depends on the context of the specific study.  Nevertheless,   Figure~\ref{fig:fp} illustrates the parameter space where misinterpretation of pulsating ULXs would contaminate the fundamental plane.

\begin{figure*}
    \centering
    \includegraphics[width=7.0in]{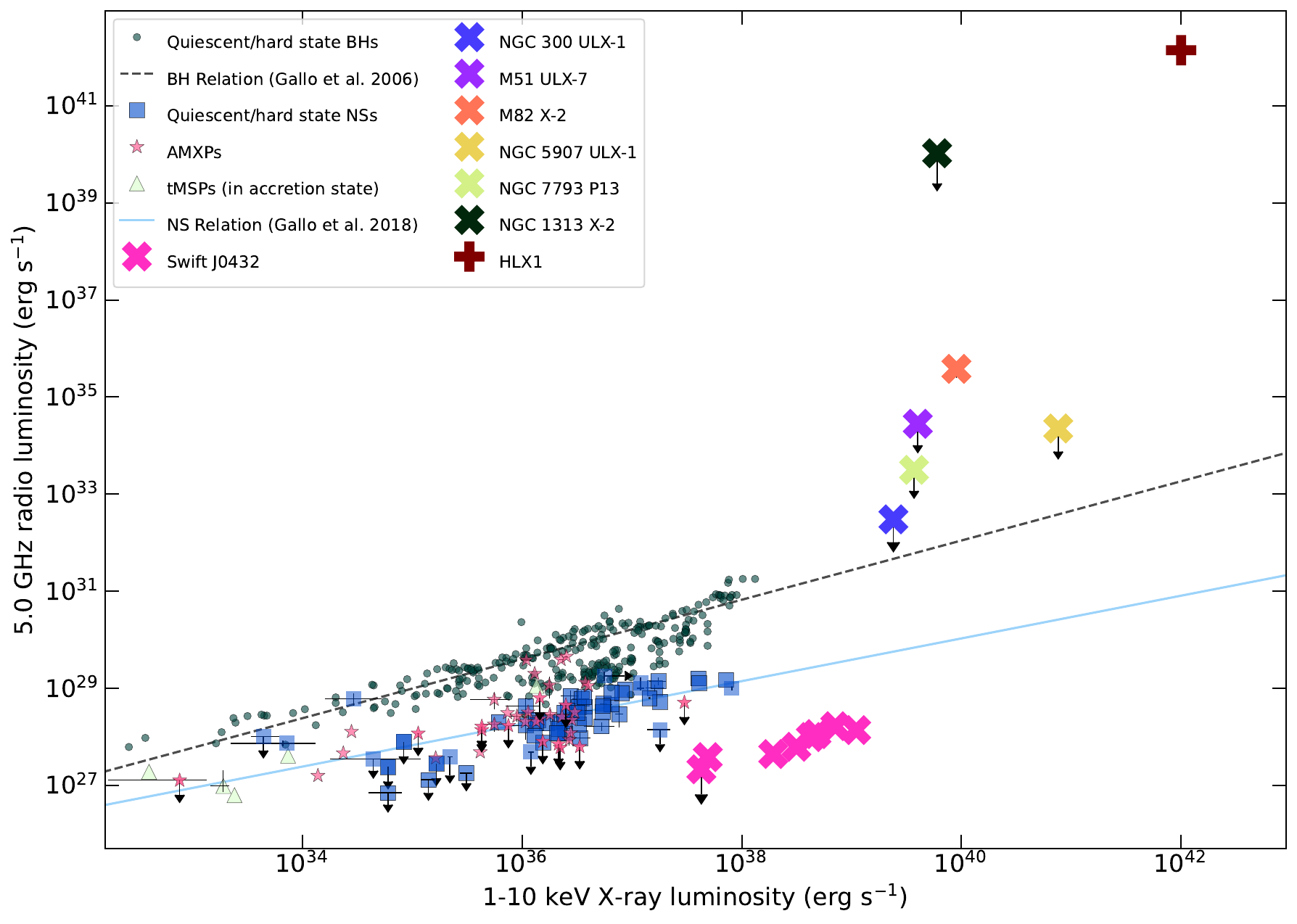}
    \caption{The radio/X-ray luminosity correlation plane for accreting compact objects adapted from \citet{Bahramian18}. Quiescent/hard state black holes are indicated by the dark green circles. We included the following neutron star populations: quiescent/hard state neutron stars as blue squares, accreting millisecond X-ray pulsars (AMXPs) and transitional millisecond pulsars (tMSPs) as pink stars and mint triangles. The neutron star ULXs in our sample are indicated by large X's. We also include HLX-1 \citep{Cseh2015} for comparison. The gray line represents the best-fit regression for black holes  from \citet{Gallo06} while the blue line represents the proposed correlation for hard state neutron stars from \citet{Gallo18}.}
    \label{fig:lrlx}
\end{figure*}

On the other hand, if one compares our sample only to other stellar-mass compact objects,  neutron star ULXs do appear distinguishable from other objects in the radio/X-ray luminosity correlation (see Figure \ref{fig:lrlx}), although this is not an appropriate comparison since stellar mass systems on the Fundamental Plane should not be above Eddington, as they would no longer be in the hard-state.  Our sample of neutron star ULXs fall into two classes in Figure \ref{fig:lrlx}, a lower X-ray luminosity source detected in radio (Swift J0243), and higher X-ray luminosity sources not detected in radio beyond the background emission. However, given the persistent ULXs are all upper limits in radio, it is possible that the true measurements may be broadly consistent with Swift J0243.

A major caveat, of course, is that we cannot compare the location of our pulsating ULXs to ULXs known to be powered by stellar-mass black holes (since it is unclear if non-pulsating ULXs are powered by black holes or neutron stars).  Nevertheless, it is worth commenting that the detected radio emission from the Galactic ULX Swift J0243  \citep{vandenEijnden18} is significantly less luminous than the radio emission from hard state neutron stars from \citet{Gallo18}. As discussed in \citet{vandenEijnden18}, this is likely due to the source's high magnetic field. 

One of the leading explanations for how neutron star ULXs can produce such a high X-ray luminosity is thanks to a high magnetic field strength \citep[e.g.][]{2018NatAs...2..312B}. If this is indeed the case, then it is not a surprise that the (more distant) neutron star ULXs were not detected in radio. 

\section{Summary and Conclusions} \label{sec:summary}
We analyzed radio observations of neutron star ULXs in conjunction with existing studies to complete the largest radio study of neutron star ULXs. We combined the radio observations with published X-ray luminosities to assess whether neutron star ULXs are likely to fill the same radio/X-ray parameter space as bona fide IMBHs. Our results are as follows:

\begin{itemize}
    \item[ 1. ]In X-ray/radio parameter space, Swift J0243 is highly dissimilar to the other pulsating ULXs, both because it is actually detected in radio (although it is also the only Galactic system), but also because it is much fainter in X-ray. This suggests that Swift J0243 may be dissimilar from the more luminous extragalactic population of neutron star ULXs.
    \item [2. ]If one were to take M82 X-2's radio counterpart at face value, it would imply a rather massive black hole. However, as reported by \citet{2013MNRAS.431.1107G} the radio counterpart is most likely a HII region near the XRB. This highlights the importance of taking background environments into consideration.
    \item [3.] NGC 300 ULX-1, NGC 7793 P13, NGC 5907 ULX-1 and M51 ULX-7 occupy a quadrant of $L_R/L_X$ where, given the intrinsic scatter in the relation, one could misappropriate the radio upper limit to infer a mass range consistent with an IMBH. We thus urge extreme caution when using $L_R/L_X$ to calculate black hole mass upper limits based on radio non detections. 
\end{itemize}

 With current facilities, we do not detect a significant radio counterpart to any extragalactic neutron star ULX. Although we are mainly presenting upper limits in this study, there are are two major conclusions to be drawn: the first is that until a larger sample of neutron star ULXs are uncovered, or we have increased capabilities from facilities such as the next generation VLA \footnote{\url{https://ngvla.nrao.edu/}} and Square Kilometre Array \footnote{\url{https://www.skao.int/}}, our study strongly suggests that, once background star formation has been properly accounted for, it is possible to use radio detections to differentiate neutron star ULXs from massive black hole ULXs. In contrast to our sample of pulsating ULXs, \citet{2015MNRAS.446.3268C}'s study of HLX-1 demonstrates that, with due caution, $L_R/L_X$ can be successfully used to interpret a radio counterpart to a ULX and provide a complementary mass estimate.

Our second conclusion is that we urge caution with the implementation of $L_R/L_X$, particularly for sources which are not detected in radio. Our results further demonstrate the incompatibility of $L_R/L_X$ and ULXs in the super Eddington state.

 \begin{acknowledgments}

The authors thank Eric Koch and Rupali Chandar for helpful discussion. KCD acknowledges support for this work
provided by NASA through the NASA Hubble Fellowship grant
HST-HF2-51528 awarded by the Space Telescope Science Institute, which is operated by the Association of Universities for Research in Astronomy, Inc., for NASA, under contract NAS5–26555. This work was partially supported by the Australian government through the Australian Research Councils Discovery Projects funding scheme (DP200102471).  
 RMP and JDP acknowledge support from the National Science Foundation under grant AST-2206123. The Australia Telescope Compact Array is part of the Australia Telescope National Facility (https://ror.org/05qajvd42) which is funded by the Australian Government for operation as a National Facility managed by CSIRO. We acknowledge the Gomeroi people as the Traditional Owners of the Observatory site. This work made use of data supplied by the UK Swift Science Data Centre at the
University of Leicester. 

 \end{acknowledgments}
\vspace{5mm}
\facilities{VLA, ATCA, Chandra, Swift}

\software{astropy \citep{Robitaille13}, CASA \citep{CASA2022}, matplotlib \citep{Hunter07}, NumPy \citep{harris20}, pandas \citep{Mckinney10}, CIAO \citep{Fruscione06}, \textsc{xspec} \citep{Arnaud96}}

\bibliography{references}{}
\bibliographystyle{aasjournal}

\end{document}